# Characteristics of Correlated Photon Pairs Generated in Ultra-compact Silicon Slow-light Photonic Crystal Waveguides

Chunle Xiong[1,*], Christelle Monat[1,2], Matthew J. Collins[1], Alex S. Clark[1], Christian Grillet[1], Graham D. Marshall[3], M. J. Steel[3], Juntao Li[4], Liam O'Faolain[4], Thomas F. Krauss[4], and Benjamin J. Eggleton[1]

[1]Centre for Ultrahigh-bandwidth Devices for Optical Systems (CUDOS), the Institute of Photonics Optical Science (IPOS), School of Physics, University of Sydney, NSW 2006, Australia
[2]Institut des Nanotechnologies de Lyon, Ecole Centrale de Lyon, 36 Avenue Guy de Collongue, 69134 Ecully, France
[3]CUDOS, MQ Photonics Research Centre, Dept. of Physics and Astronomy, Macquarie University, NSW 2109, Australia
[4]School of Physics and Astronomy, University of St Andrews, Fife, KY16 9SS, UK

*Abstract*—We report the characterization of correlated photon pairs generated in dispersion-engineered silicon slow-light photonic crystal waveguides pumped by picosecond pulses. We found that taking advantage of the 15 nm flat-band slow-light window ($v_g \sim c/30$) the bandwidth for correlated photon-pair generation in 96 and 196 µm long waveguides was at least 11.2 nm; while a 396 µm long waveguide reduced the bandwidth to 8 nm (only half of the slow-light bandwidth due to the increased impact of phase matching in a longer waveguide). The key metrics for a photon-pair source: coincidence to accidental ratio (CAR) and pair brightness were measured to be a maximum 33 at a pair generation rate of 0.004 pair per pulse in a 196 µm long waveguide. Within the measurement errors the maximum CAR achieved in 96, 196 and 396 µm long waveguides is constant. The noise analysis shows that detector dark counts, leaked pump light, linear and nonlinear losses, multiple pair generation and detector jitter are the limiting factors to the CAR performance of the sources.

## I. INTRODUCTION

Correlated photon pairs at around 1550 nm are one of the most important elements for implementing photonic quantum information technologies that are compatible with telecommunication systems. For example, a photon-pair source can be used as a heralded single-photon source or as a polarization or time-bin entangled source [1]–[17]. This has motivated the development of photon-pair generation in silica optical fibers [1]–[7] with a view towards all-fiber quantum information systems [8]. In particular, for integrated quantum information processing, photon-pair generation has been demonstrated in a number of photonic chip platforms, such as crystalline and amorphous silicon nanowires [9]–[14], periodically poled $LiNbO_3$ (PPLN) [15] and $LiTO_3$ (PPLT) [16] waveguides, and chalcogenide waveguides [17].

All of the schemes in [1]–[17] produce photon pairs through spontaneous nonlinear processes that are stochastic, and one of the major challenges in the field is to deterministically generate photons with negligible contamination from higher photon number states. The reliable generation of single photons on demand will require many photon-pair sources pumped with sufficiently low power to preclude higher photon number states and multiplexed to form a parallel architecture [18], [19]. To achieve the ultimate goal of integrating nonlinear photon-pair sources and linear photonic circuits [20] on a monolithic chip, we eventually need to critically reduce the size of each individual pair generation unit so that hundreds or even thousands of them can be fitted onto a small integrated photonic device.

In the regime of nonlinear pair generation, the flux of generated photon pairs grows with the effective nonlinear interaction strength $\gamma PL$, where $P$ is the pump power, $L$ the device length, and $\gamma$ measures the strength of the optical nonlinearity that depends on the material and geometry of the device. For a given power consumption, the required path length of a nonlinear device can be significantly reduced if we can enhance the device nonlinearity. For example, in previous photon-pair generation experiments, the typical length of silica fibers was in the range 1–300 m due to the low silica nonlinearity and the weak optical confinement provided by the fibers [1]–[8], while the length of PPLN, PPLT and chalcogenide waveguides was reduced to a few centimeters [15]–[17]. The use of highly nonlinear silicon nanowires, where light is tightly confined down to the sub-micrometer scale, has been shown to enable decreases in the device length down to approximately 1 cm [9]–[14]. Very recently, our team has reported that photon pairs can be efficiently generated in a much more compact 96 µm long silicon photonic crystal (PhC) waveguide [21]. This is two orders of magnitude shorter than the shortest silicon nanowires previously reported, and is possible due to the effective enhancement in nonlinearity gained by dispersion engineering and slow-light propagation in these periodic structures [22]–[24]. However, the maximum CAR in [21] was limited to 13, partly due to insufficient pump suppression. To date, the characteristics of photon-pair generation in this platform and the associated challenges and limitations have not been presented in full detail.

*Email: chunle@physics.usyd.edu.au

Following our initial report in [21], we present detailed characterization of these slow-light based silicon photon-pair sources and show, in a thorough study, how to improve their performance. We report the frequency detuning and device length dependency of pair generation through spontaneous four-wave mixing (SFWM) in silicon slow-light PhC waveguides with a group index close to 30. While the SFWM efficiency ideally scales as $(\gamma PL)^2$ in lossless waveguides and with all three signals perfectly phase matched [22], deviations of the pair generation rate from this trend are observed due to the slow-light dependence of the propagation loss, nonlinear loss and the increasing impact of phase matching for longer waveguides. We show that in 96 and 196 µm long waveguides, the SFWM bandwidth is at least 1.4 THz (~11.2 nm), which almost spans the 15 nm bandwidth of the flat-band slow-light window, where dispersion is engineered and significantly reduced ($\beta_2 \approx 1 \times 10^{-21}$ s$^2$/m) [22], [23]. The 396 µm long waveguide, however, has a reduced bandwidth of 1 THz (~8 nm), which is only half of the flat-band slow-light window due to poorer phase matching after a longer propagation distance. These observations are consistent with the four-wave mixing bandwidth measurement in the stimulated regime [22], [23]. In our experiment, a maximum CAR of 33 was achieved in the 196 µm waveguide. We also confirmed that the maximum CAR achieved in the 96, 196 and 396 µm long waveguides and obtained at different pump power was constant within the measurement errors. We finally examined the origin of noise by comparing the single and coincidence counts and found that besides detector dark counts, the insufficient pump suppression, linear and nonlinear losses, multiple pair generation and detector jitter were the limiting factors to the CAR performance of the pair sources.

## II. EXPERIMENTAL SETUP

The generation of correlated photon pairs through SFWM is illustrated in Fig. 1(a). A coherent pulse of light enters the PhC waveguide, where two photons from the pump are converted to signal and idler photons of higher and lower frequencies respectively to form a quantum correlated state. The 96, 196 and 396 µm silicon slow-light devices used in this study were fabricated on the same silicon-on-insulator wafer comprising a 220 nm silicon layer on 2 µm of silica using electron beam lithography and reactive ion etching [21]–[23]. The PhC waveguides were created from a triangular lattice of air holes etched in a suspended silicon membrane with a row of holes missing along the ΓK direction. The two rows adjacent to the waveguide center were laterally shifted to engineer the waveguide dispersion such that it exhibits a group index of 30±10% with low dispersion ($\beta_2 \approx 1 \times 10^{-21}$ s$^2$/m) and moderate loss across a 15 nm window [24]. Silicon access waveguides, including inverse tapers terminated by wide polymer waveguides, were added to the input and output of the PhC region to improve coupling efficiency [23]. Figure 1(b) shows the measured group index and transmission of the 96 µm PhC waveguide. We can see that at 1555 nm, the total insertion loss of the 96 µm device is about 8 dB. The 196 and 396 µm devices have similar group index curves, but the insertion losses at 1555 nm are about 9 and 11 dB, respectively, due to higher propagation loss in longer waveguides. The input and output coupling losses between the waveguides and off-chip components are assumed to be the same and estimated to be 3.5 dB per facet for the three PhC waveguides.

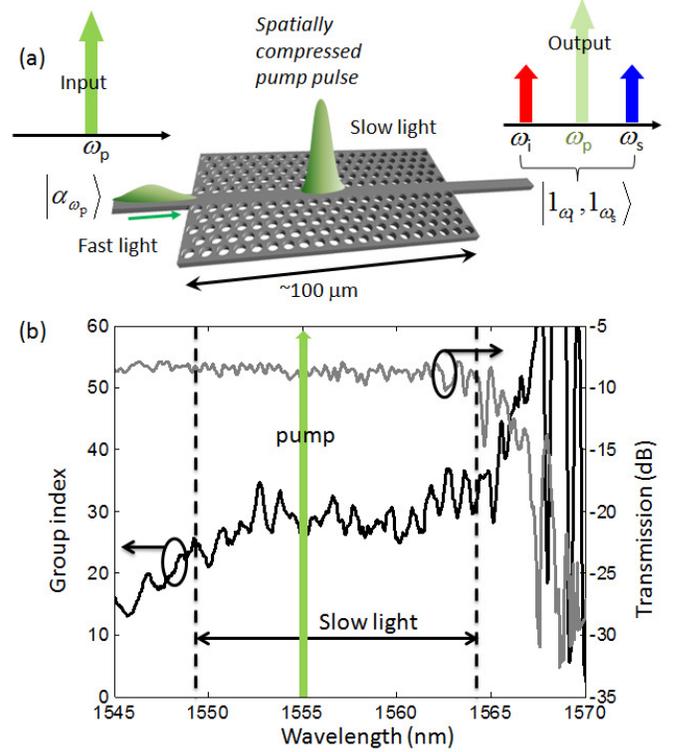

Fig. 1. (Color online) (a) Schematic of SFWM in a silicon slow-light PhC waveguide. (b) Group index and total transmission of light in the 96 µm long silicon PhC waveguide. The window between dotted lines with a group index of 30±10% and slightly increased loss defines the flat-band slow-light window. The green arrow in (b) indicates the pump wavelength.

Figure 2 shows our experimental setup. A mode-locked fiber laser delivering a 10 MHz pulse train at 1554.9 nm was filtered using a 0.5 nm bandpass filter (BPF) and a 980/1550 wavelength division multiplexing (WDM) coupler to eliminate the out of pump band noise. The full width at half maximum (FWHM) length of the pulses after filtering was measured to be 10 ps. The pump beam was TE polarized with a polarization controller and a polarizer before coupling to the chip using a lensed fiber. A fiber attenuator was used to control the input pump power. Signal and idler photons generated at higher and lower frequency than the pump, respectively, were directed to a circulator and a fiber Bragg grating (FBG) to suppress pump photons. They were then post-selected in frequency and separated using an arrayed waveguide grating (AWG) and further filtered using 0.5 nm BPFs before being sent to InGaAs/InP single-photon detectors (SPDs, id-Quantique 201). The AWG had 40 channels with frequency spacing of 100 GHz and channel FWHM of 50 GHz. The wavelength tunable BPFs and the AWG allowed us to post-select the frequency detuning of the photon pairs in a range of $\Delta f = \pm 0.7$ THz from the pump frequency. The SPDs operated in the gated mode. Because the maximum gate frequency of the SPDs is 8 MHz, we employed a delay generator (DG1, Stanford Research Systems DG645) to select alternate RF pulses from the laser to trigger SPD1 with an effective gate frequency of 5 MHz. The detection output signals from SPD1 were appropriately delayed by DG2 and used to trigger SPD2 for coincidence and accidental coincidence measurements, successively. The nominal gate width for both SPDs was set at 2.5 ns, which is the shortest possible duration for single-photon detection with minimum dark count. The detection probability for both SPDs was set to 10%, at which the effective gate width is 0.5 ns due to the imperfect rectangle shape of the gate pulse, and the dark counts were measured to be $4\times10^{-5}$ and $2\times10^{-5}$ per gate for idler and signal detectors, respectively. Taking into account the waveguide output coupling loss, the circulator, FBG, AWG and BPFs insertion losses, and the detection efficiency, the total loss of each channel was estimated to be 22 dB. To improve measurement statistics, we acquired counts for 30 minutes at < 0.4 W coupled peak power and 5 minutes at higher power.

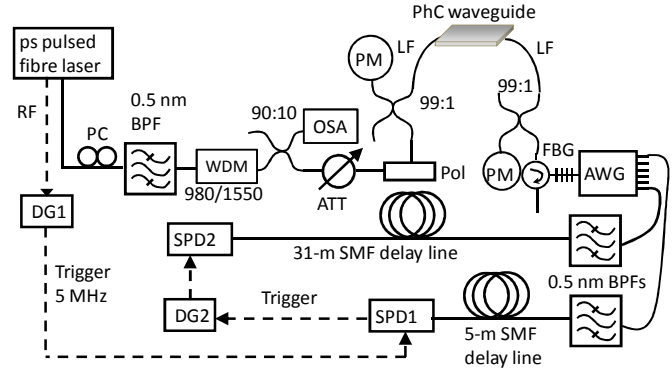

Fig. 2. Experimental setup. PC: polarization controller, BPF: bandpass filter, ATT: attenuator, PM: power meter, LF: lensed fiber, FBG: fiber Bragg grating, AWG: arrayed waveguide grating, SMF: single-mode fiber, SPD: single-photon detector, DG: delay generator, RF: radio frequency.

## III. EXPERIMENTAL RESULTS

### A. Influence of Pump Leakage on CAR

In our previous report [21], one of the main limiting factors for maximum CAR was pump leakage. In this study, we examined the effect of pump suppression using a FBG. The FWHM bandwidth of the FBG was 0.5 nm and the central wavelength was tunable from 1545 to 1555 nm. We post-selected the signal and idler photons generated from the 96 µm waveguide at a fixed frequency detuning of $|\Delta f|=0.7$ THz (i.e. 5.6 nm). We measured the coincidence ($C_{raw}$) and accidental coincidence ($A$) at different pump powers with the FBG central wavelength tuned to the pump wavelength and offset from it by 1 nm, respectively. The CAR was calculated as CAR=$C/A$ with $C=C_{raw}-A$ as the net coincidence. A CAR>0 shows the existence of a time correlation, and a higher CAR indicates a stronger temporal correlation between the signal and idler photons. The CAR as a function of coupled peak power for both cases is shown in Fig. 3. The coupled peak power used in the plot is calculated from the measured input power in the launch fiber and estimated coupling loss at the chip end-facet. It can be seen that in both cases, the CAR increases with the power until it reaches a maximum and then starts to decay. The existence of a peak is the result of competition between actual photon pairs and noise induced by detector dark counts and after-pulsing, leaked pump photons, linear and nonlinear losses, and multiple pair generation (Raman noise is absent in the silicon platform [11]). In the ideal case of no extra noise other than multiple pair generation, the CAR of a nonlinear photon pair source should decrease monotonically as the pump power increases [11]. In practice, however, detector dark counts and pump leakage noise are dominant at low pump power. As the pump power increases, pair generation overcomes these sources of noise and the CAR increases until detector after-pulsing, nonlinear loss and multiple pair effects become dominant (in our study this occurs at powers exceeding 0.2 W). The optimum CAR was enhanced to 27 when the FBG was aligned with the pump wavelength to block more pump photons. In fact, while the FBG used in the experiment has the same FWHM bandwidth as the pump, it has a narrower -20 dB width than the pump, so the measured pump suppression was only 12 dB rather than the expected 30 dB. A further enhanced CAR would be expected if a much broader FWHM bandwidth FBG were to be used to more effectively block the pump. From the plot of the net coincidence as a function of coupled peak power in Fig. 3, we can see that the net coincidence starts to deviate from the quadratic dependence at 0.5 W, where the nonlinear loss occurs. At further higher power, the nonlinear loss and multiple pair generation become the dominant sources of noise affecting the CAR, so the pump leakage has less effect and the CAR for both cases converges.

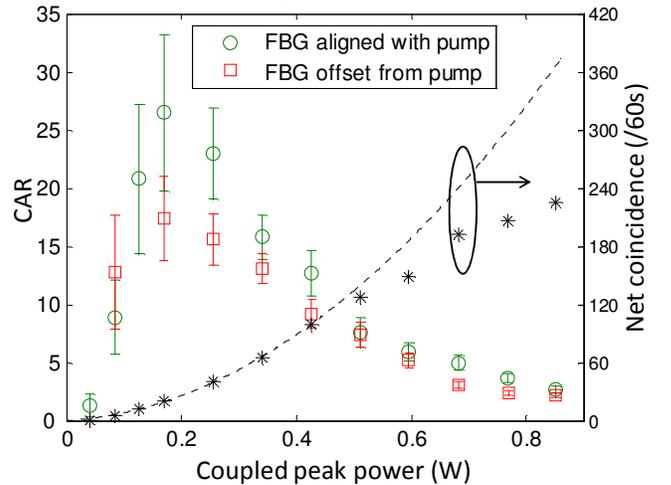

Fig. 3. (Color online) The CAR (left axis) and net coincidence (right axis, stars) as a function of coupled peak power for the 96 µm PhC waveguide. Poissonian error bars are used for the CAR plot. The dotted line is a quadratic fit for net coincidence.

### B. Influence of Photon-Pair Frequency Detuning

Next we investigated the influence of frequency detuning $|\Delta f|$ between the pump and photon pairs. At small $|\Delta f|$ pairs will become swamped by leaked pump light due to the limited performance of the optical components such as the FBG, AWG and BPFs, so $|\Delta f|$ should be large enough to avoid this.

On the other hand however, |Δ$f$| is constrained to the bandwidth of the slow-light regime and SFWM, as well as the tuning range of AWG and BPFs.

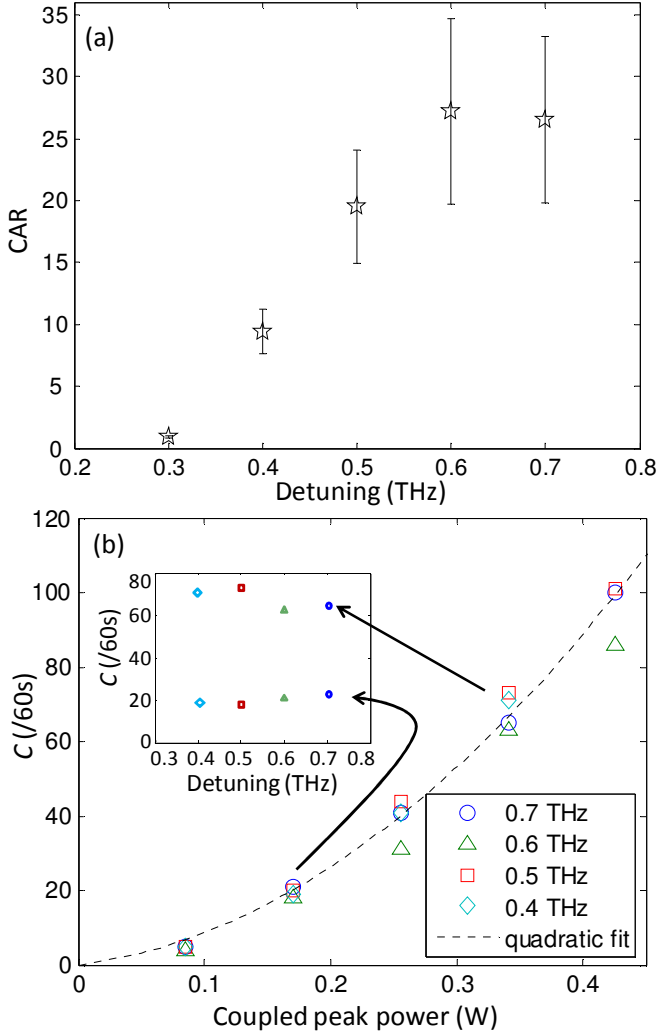

Fig. 4. (Color online) (a) The CAR at fixed coupled peak power of 0.17 W as a function of frequency detuning for the 96 µm waveguide. Poissonian error bars are used. (b) The net coincidence count as a function of coupled peak power at different detuning. Blue circles, green triangles, red squares and cyan diamonds are for 0.7, 0.6, 0.5 and 0.4 THz, respectively. The dotted black line shows a quadratic fit. The inset shows the net coincidence count as a function of detuning at 0.17 and 0.34 W coupled peak power. Error bars are small compared with the data points and are not shown.

Figure 4(a) shows the CAR at fixed coupled peak power of 0.17 W as a function of |Δ$f$| for the 96 µm waveguide. For |Δ$f$|=0.6 and 0.7 THz, the CAR is similar and relatively high (unfortunately we were unable to tune beyond 0.7 THz due to limitations in the AWG bandwidth). The CAR drops quickly for |Δ$f$|<0.6 THz suggesting that, although the raw detector count rates increase as the measurement of photon-pair frequencies approach the pump, more pump photons leak into the signal and idler channels thereby degrading the CAR. To further confirm this, we examined the net coincidence count corrected for accidentals ($C=C_{raw}–A$, which compensates for leaked pump photons) at different detunings, and plotted it as a function of coupled peak power in Fig. 4(b). The inset of Fig. 4(b) shows the net coincidence count as a function of frequency detuning at two pump power levels. It can be seen that $C$ is independent of detuning for given pump power and $C$ increases quadratically with the coupled peak power, which is a signature of SFWM in silicon platforms when they are pumped at low power to avoid nonlinear loss. These observations confirm that the frequency detuning apparent dependence of the CAR is caused by increasing pump leakage for smaller detuning. This pump leakage degrades the CAR through increasing the accidental coincidence, while the SFWM efficiency remains relatively constant within the slow-light window, as attested by the steady net coincidence counts. The uniformity agrees with the stimulated measurements using the pump-probe technique in the classical regime [22], where the conversion effiency was found to be flat in the range of |Δ$f$|≤1 THz (i.e. within the whole flat-band slow-light window).

### C. Waveguide Length Dependence

Next we investigated the PhC waveguide length dependence of the correlated photon-pair generation, which might yield the existence of an optimum length in terms of efficiency, bandwidth and CAR of the source. Figure 5(a) shows the measured net coincidence count as a function of coupled peak power for waveguide lengths of 96 (blue circles), 196 (green squares) and 396 (red triangles) µm. The dotted lines show quadratic fits, and as expected, all three curves depend on power quadratically when nonlinear loss is absent (i.e. at low power) with the nonlinear loss induced deviation from the quadratic dependence occurring at lower power in the two longer waveguides. In addition, more pairs are generated in longer waveguides than in shorter ones when the power is the same. However, we can notice from the inset of Fig. 5(a) that the net coincidence counts do not increase quadratically with device length because of higher propagation loss and probably poorer phase matching in longer waveguides.

It should be mentioned that in Fig. 5(a), the data for the 96 and 196 µm waveguides were measured at |Δ$f$|=0.7 THz but the data for the 396 µm waveguide were taken at |Δ$f$|=0.5 THz. The detuning providing the maximum CAR, as studied in section B, is lower in the longer waveguide than that in the shorter ones because the associated SFWM bandwidth is narrower due to the increased impact of phase matching [22], [23]. To illustrate this, we plot, as in the Fig. 4(b) inset, the net coincidence count as a function of detuning at 0.17 W coupled peak power for both 96 and 396 µm PhC waveguides in Fig. 5(b). While the SFWM efficiency was found relatively independent on the detuning in the 96 µm waveguide, it dropped significantly with increased detuning in the 396 µm waveguide. This bandwidth measurement at the single-photon level agrees with that using the classical stimulated FWM measurement reported in [22] and [23]. An explanation for this is that even though we tried to make PhC waveguides with near-zero group velocity dispersion over the whole slow-light window, the residual non-zero dispersion ($β_2≈1×10^{-21}$ s$^2$/m) degrades phase matching at larger detuning in longer waveguides [22], [23].

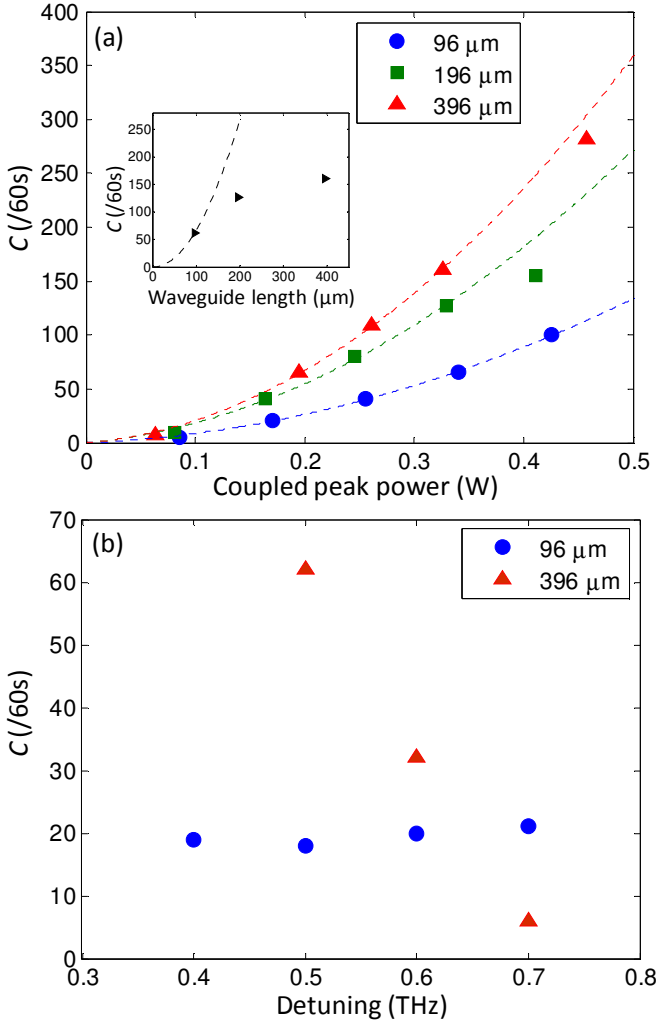

that the maximum CAR does not have a strong dependence on the waveguide length if we take into account the measurement errors. The power at which this maximum CAR value was achieved only slightly decreased from 0.17 W to 0.13 W for a four time increase in the waveguide length, making the optimum *PL* product higher for the longer waveguide. This is understood as the consequence of balance between pair generation efficiency, propagation loss and pump leakage. For example, to generate a given number of photon pairs, we need less pump power for a longer waveguide so that we have less pump leakage, but the associated higher propagation loss counteracts this benefit, resulting in a similar maximum CAR performance as in a shorter waveguide. In conclusion, there is no real optimum length of the PhC based photon-pair source, as far as improving the CAR is concerned, but if compactness is not an issue, increasing the length still slightly relaxes the peak power needed to achieve the highest CAR performance.

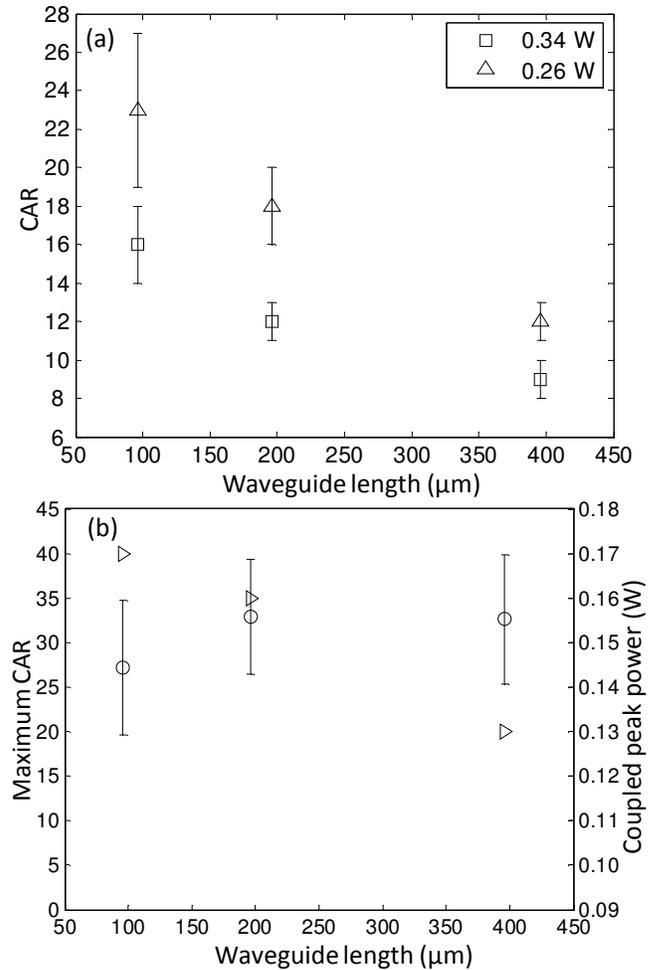

Fig. 5. (Color online) The measured net coincidence count as a function of (a) coupled peak power for waveguide lengths of 96 (blue circles), 196 (green squares) and 396 (red triangles) µm, and (b) frequency detuning for wavelength lengths of 96 (blue circles) and 396 (red triangles) µm at coupled peak power of 0.17 W. The inset in (a) shows the net coincidence count as a function of waveguide length at the same coupled peak power of 0.34 W. The dotted lines in (a) are quadratic fit.

We next investigated the dependence of CAR on PhC waveguide length, at two different coupled peak powers of 0.34 and 0.26 W (see Fig. 6(a)). For each pump power, the CAR decreases with the device length. This is explained by the higher probability of multiple pair generation in a longer waveguide, and is consistent with the increasing SFWM efficiency for longer waveguides inferred from Fig. 5(a). A mathematical fit shows that the CAR change follows the rule of $L^{-0.5}$, which is in contrast to the ideal $L^{-2}$ dependence in a statistical model [11] due to the influence of propagation loss, detector dark counts and pump leakage.

To find the maximum possible CAR, we performed the CAR measurement for the three waveguides at different powers and plotted the maximum achieved CAR as a function of waveguide length and the corresponding coupled peak power in Fig. 6(b). We found that the optimum CAR was 33 at a pair generation rate of 0.004 pair per pulse and detuning of 0.7 THz in the 196 µm long waveguide. Yet, Fig. 6(b) shows

Fig. 6. (a) The CAR as a function of PhC waveguide length at coupled peak power of 0.34 (squares) and 0.26 (triangles) W. (b) The maximum CAR (left axis, circles) as a function of waveguide length and the corresponding coupled peak power used to achieve the maximum CAR (right axis, triangles). Poissonian error bars are used for the CAR.

### D. Noise analysis: Single and Coincidence Counts

At the power level where the maximum CAR of 33 was achieved, the measured net coincidence count in 30 minutes was 1217. After taking into account the 22 dB loss in each channel and the trigger rate, we can infer the actual pair generation rate at the output end of the 196 µm PhC waveguide to be 0.004 pair per pulse. According to the statistical model in [11], for a pair generation rate $\mu$ (average number of pairs per pulse), in the absence of noise, the ideal CAR is given by $1/\mu=1/0.004=250$. If the detector dark counts and channel linear losses are included in the model [11], the CAR falls to 51. To understand the discrepancy between the measured and predicted maximum CAR, we investigated and compared the single and coincidence counts.

As shown in the experimental setup in Fig. 2, we measured net single counts ($S=S_{raw}-D$) using SPD1 and net coincidence counts ($C$) using SPD2, where $S_{raw}$ and $D$ represent the measured raw single counts and dark counts, respectively. Then we can estimate single ($\mu_1$) and pair ($\mu_2$) generation rates at the output end of the waveguides using $\mu_1=S/(t\eta R)$ and $\mu_2=C/(t\eta^2 R)$, where $t$ is the integration time for collecting counts, $R$ is the trigger rate of SPD1 and $\eta$ is the photon collection efficiency in each channel. As $\mu_2$ is calculated from a correlated coincidence, all uncorrelated noise is excluded. $\mu_1$, however, may include leaked pump photons, one of the pair photons with the other lost due to linear and nonlinear propagation losses on chip, and photons from different pairs. Therefore the ratio $\mu_1/\mu_2$ and its dependence on the pump power will help us to understand the noise source in the system.

In the silicon nanowire experiment of Harada *et al* [11], it was reported that at the pair generation rate of 0.001 pair per pulse, $\mu_1/\mu_2=1$ at $|\Delta f|\geq 0.6$ THz, which means that the effect of pump leakage, multiple pair generation, linear and nonlinear losses under their experimental condition is negligible. In our study, we plot the ratio $\mu_1/\mu_2$ as a function of coupled peak power at different detuning for all three waveguides in Fig. 7. There are three features on the plots. First, the ratio goes up when the frequency detuning is smaller for the 96 µm device, but does not change significantly for the two longer devices. This can be understood as a result of higher relative pump leakage at smaller detuning in the 96 µm device. In the two longer devices, however, the pump leakage effect is relatively weaker because of the higher pair generation rate (see Fig 5(a)). Second, all plots appear to have a minimum ratio at a certain power $P_m$. The ratio increases at a lower power than $P_m$ due to relatively higher number of leaked pump photons with respect to the generated pairs; this phenomenon is more obvious in a shorter waveguide for the same reasons as the first feature. The increase of $\mu_1/\mu_2$ at a higher power than $P_m$ can be explained by the increasing nonlinear propagation loss in the PhC waveguide, causing one of the pair photons to be absorbed and also due to multiple pair generation. Third, the minimum ratio is about 2.5, and roughly the same for all waveguide lengths. Compared with the ideal $\mu_1/\mu_2=1$, we attribute this discrepancy to linear propagation loss, pump leakage and the detector jitter. The effects of linear propagation loss and pump leakage are straightforward. We will focus on understanding the effect of detector jitter.

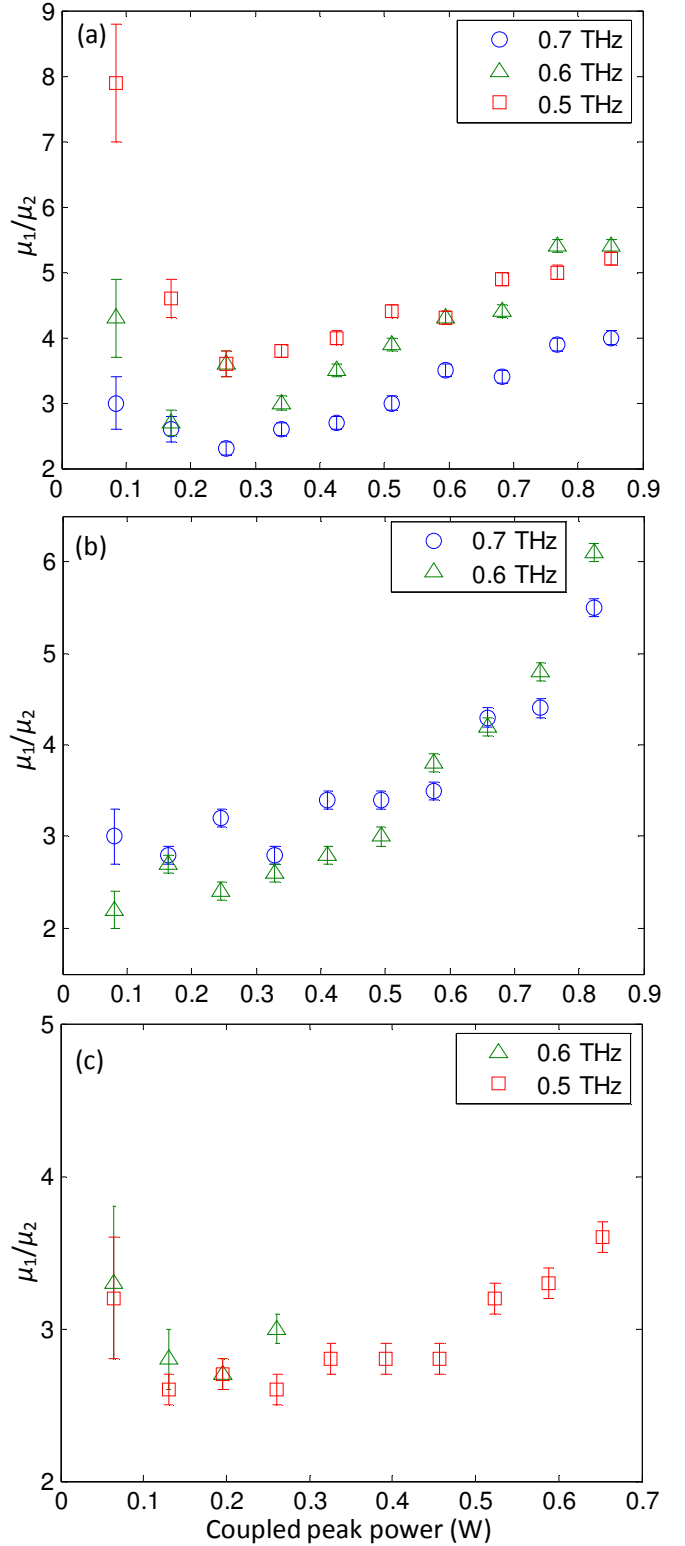

Fig. 7. (Color on line) The measured $\mu_1/\mu_2$ as a function of coupled peak power at different detuning for waveguide lengths of (a) 96, (b) 196 and (c) 396 µm. The blue circles, green triangles and red squares are for frequency detuning of 0.7, 0.6 and 0.5 THz, respectively.

The generated photon pairs that are not lost across the idler and signal channels arrive at the detectors at a well defined time; however, the opening of the detector gate may not be exactly synchronized with the photon arrival time because of the detector jitter. Only when the effective gate width of the detector covers the jitter induced uncertainty, is the detector able to detect all arriving photons. This is the case for the single measurement at SPD1, because the jitter induced uncertainty of the time at which the gate is opened on SPD1 was measured to be 0.7 ns, only slightly larger than the effective gate width of 0.5 ns. For the coincidence measurement at SPD2, however, the jitter induced uncertainty of the SPD2 gate opening time was measured to be 1.1 ns, much bigger than the 0.5 ns effective gate width, which may significantly lower the number of detected photons out of the photons reaching the SPD2 detector. The SPD2 jitter induced uncertainty was bigger than the SPD1 one because SPD1 was triggered by a low-jitter RF signal from the laser whereas SPD2 was triggered by the high-jitter output from SPD1. Note that if we use a wider gate width on SPD2, i.e. closer to the associated jitter uncertainty, the increased dark count would degrade the CAR. To overcome this issue, our future measurement setup will have to trigger both detectors using the RF signal from the pump laser, enabling us to collect all photons within the 0.5 ns effective gate window, and will be completed by a time interval analyzer to record the coincidence [11].

IV. CONCLUSION

In conclusion, we have systematically characterized an ultra-compact silicon slow-light PhC waveguide platform based correlated photon-pair source. We have confirmed that the bandwidth of the pair generation using SFWM in a 96 and 196 µm long device is at least 1.4 THz (11.2 nm), which almost makes use of the full 15 nm wide slow-light window of the device. The bandwidth is much narrower in a 396 µm long device because of poorer phase matching. We have found that the maximum CAR does not have a strong dependence on the device length due to the balance between multiple noise factors such as pump leakage and propagation loss, and pair generation efficiency. A maximum CAR of 33 was achieved at pair generation rate of 0.004 pair per pulse. The maximum CAR is mainly affected by detector dark count, insufficient pump suppression and linear propagation loss. The investigation of single and coincidence counts shows that the nonlinear loss and multiple pair generation do exist at high power in this silicon platform. Some non-ideality also arises from our setup arrangement, rather than the source itself. In particular, to overcome the detector jitter issue occurring in our measurement system, a time interval analyzer may have to be used in the future.

As ultra-compact photon sources are the key element for emerging quantum technologies, this study provides a comprehensive understanding to this important platform and promotes a scalable approach for quantum information processing on-chip.

REFERENCES

[1] X. Li, J. Chen, P. Voss, J. Sharping, and P. Kumar, "All-fiber photon-pair source for quantum communications: improved generation of correlated photons," *Opt. Express*, vol. 12, no. 16, pp. 3737–3744, Aug. 2004.
[2] H. Takesue and K. Inoue, "1.5-µm band quantum-correlated photon pair generation in dispersion-shifted fiber: suppression of noise photons by cooling fiber," *Opt. Express*, vol. 13, no. 20, pp. 7832–7839, Oct. 2005.
[3] S. D. Dyer, B. Baek, and S. W. Nam, "High-brightness, low-noise, all-fiber photon pair source," *Opt. Express*, vol. 17, no. 12, pp. 10290–10297, Jun. 2009.
[4] H. Takesue, "1.5 µm band Hong-Ou-Mandel experiment using photon pairs generated in two independent dispersion shifted fibers," *Appl. Phys. Lett.*, vol. 90, pp. 204101-1–3, May 2007.
[5] A. R. McMillan, J. Fulconis, M. Halder, C. Xiong, J. G. Rarity and W. J. Wadsworth, "Narrowband high-fidelity all-fibre source of heralded single photons at 1570 nm," *Opt. Express*, vol. 17, no. 8, pp. 6156–6165, Apr. 2009.
[6] X. Li, P. L. Voss, J. E. Sharping, and P. Kumar, "Optical-fiber source of polarization-entangled photons in the 1550 nm telecom band," *Phys. Rev. Lett.*, vol. 94, pp. 053601-1–4, Feb. 2005.
[7] H. Takesue and K. Inoue, "Generation of 1.5-µm band time-bin entanglement using spontaneous fiber four-wave mixing and planar light-wave circuit interferometers," *Phys. Rev. A*, vol. 72, pp. 041804(R)-1–4, Oct. 2005.
[8] J. Chen, J. B. Altepeter, M. Medic, K. F. Lee, B. Gokden, R. H. Hadfield, S. W. Nam, and P. Kumar, "Demonstration of a quantum controlled-NOT gate in the telecommunications band," Phys. Rev. Lett., vol. 100, pp.133603-1–4, Apr. 2008.
[9] J. E. Sharping, K. F. Lee, M. A. Foster, A. C. Turner, B. S. Schmidt, M. Lipson, A. L. Gaeta, and P. Kumar, "Generation of correlated photons in nanoscale silicon waveguides," *Opt. Express*, vol. 14, no. 25, pp. 12388–12393, Dec. 2006.
[10] S. Clemmen, K. Phan Huy, W. Bogaerts, R. G. Baets, Ph. Emplit and S. Massar, "Continuous wave photon pair generation in silicon-on-insulator waveguides and ring resonators," *Opt. Express*, vol. 17, no. 19, pp. 16558-16570, Sep. 2009.
[11] K. Harada, H. Takesue, H. Fukuda, T. Tsuchizawa, T. Watanabe, K. Yamada, Y. Tokura, and S. Itabashi, "Frequency and polarization characteristics of correlated photon-pair generation using a silicon wire waveguide," *IEEE J. Sel. Top. Quantum Electron.*, vol. 16, no. 1, pp. 325–331, Feb. 2010.
[12] H. Takesue, H. Fukuda, T. Tsuchizawa, T. Watanabe, K. Yamada, Y. Tokura, and S. Itabashi, "Generation of polarization entangled photon pairs using silicon wire waveguide," *Opt. Express*, vol. 16, no. 8, pp. 5721–5727, Apr. 2008.
[13] K. Harada, H. Takesue, H. Fukuda, T. Tsuchizawa, T. Watanabe, K. Yamada, Y. Tokura, S. Itabashi, "Generation of high-purity entangled photon pairs using silicon wire waveguide," *Opt. Express*, vol. 16, no. 25, pp. 20368–20373, Dec. 2008.
[14] S. Clemmen, A. Perret, S. K. Selvaraja, W. Bogaerts, D. van Thourhout, R. Baets, Ph. Emplit, and S. Massar, "Generation of correlated photons in hydrogenated amorphous-silicon waveguides," *Opt. Lett.*, vol. 35, no. 20, pp. 3483–3485, Oct. 2010.
[15] M. Hunault, H. Takesue, O. Tadanaga, Y. Nishida, and M. Asobe, "Generation of time-bin entangled photon pairs by cascaded second-order nonlinearity in a single periodically poled $LiNbO_3$ waveguide," *Opt. Lett.*, vol. 35, no. 8, pp. 1239–1241, Apr. 2010.
[16] M. Lobino, G. D. Marshall, C. Xiong, A. S. Clark, D. Bonneau, C. M. Natarajan, M. G. Tanner, R. H. Hadfield, S. N. Dorenbos, T. Zijlstra, V. Zwiller, M. Marangoni, R. Ramponi, M. G. Thompson, B. J. Eggleton, and J. L. O'Brien, "Correlated photon-pair generation in a periodically poled MgO doped stoichiometric lithium tantalate reverse proton exchanged waveguide," *Appl. Phys. Lett.*, vol. 99, 081110-1–4, Aug. (2011).
[17] C. Xiong, G. D. Marshall, A. Peruzzo, M. Lobino, A. S. Clark, D.-Y. Choi, S. J. Madden, C. M. Natarajan, M. G. Tanner, R. H. Hadfield, S. N. Dorenbos, T. Zijlstra, V. Zwiller, M. G. Thompson, J. G. Rarity, M. J. Steel, B. Luther-Davies, B. J. Eggleton, and J. L. O'Brien, "Generation of correlated photon pairs in a chalcogenide $As_2S_3$ waveguide, " *Appl. Phys. Lett.*, vol. 98, pp. 051101-1–4, Jan. 2011.
[18] A. L. Migdall, D. Branning, and S. Castelletto, "Tailoring single-photon and multiphoton probabilities of a single-photon on-demand source," *Phys. Rev. A*, vol. 66, pp. 053805-1–4, Nov. 2002.
[19] X. Ma, S. Zotter, J. Kofler, T. Jennewein, and A. Zeilinger, "Experimental generation of single photons via active multiplexing," *Phys. Rev. A*, vol. 83, pp. 043814-1–8, Apr. 2011.

[20] A. Politi, M. J. Cryan, J. G. Rarity, S. Yu, J. L. O'Brien, "Silica-on-silicon waveguide quantum circuits," *Science*, vol. 320, pp. 646–649, May 2008.
[21] C. Xiong, C. Monat, A. S. Clark, C. Grillet, G. D. Marshall, M. J. Steel, J. Li, L. O'Faolain, T. F. Krauss, J. G. Rarity, and B. J. Eggleton, "Slow-light enhanced correlated photon-pair generation in a silicon photonic crystal waveguide," *Opt. Lett.* 36, no. 17, pp. 3413–3415, Sept. 2011.
[22] C. Monat, M. Ebnali-Heidari, C. Grillet, B. Corcoran, B. J. Eggleton, T. P. White, L. O'Faolain, J. Li, and T. F. Krauss, "Four-wave mixing in slow light engineered silicon photonic crystal waveguides," *Opt. Express*, vol. 18, no. 22, pp. 22915–22927, Oct. 2010.
[23] J. Li, L. O'Faolain, I. H. Rey, and T. F. Krauss, "Four-wave mixing in photonic crystal waveguides: slow light enhancement and limitations," *Opt. Express*, vol. 19, no. 5, pp. 4458–4463, Feb. 2011.
[24] J. Li, T. P. White, L. O'Faolain, A. Gomez-Iglesias, and T. F. Krauss, "Systematic design of flat band slow light in photonic crystal waveguides," *Opt. Express*, vol. 16, no. 9, pp. 6227–6232, Apr. 2008.